# A Comparison of Various Electricity Tariff Price Forecasting Techniques in Turkey and Identifying the Impact of Time Series Periods

Benli T.O. [a], Sengul H. [b]

[a] *Graduate School of Clean and Renewable Energies, Hacettepe University, P.O. Box 06800 Beytepe, Ankara, Turkey*
[b] *Environmental Engineering Department, Hacettepe University, P.O. Box 06800 Beytepe, Ankara, Turkey*

ARTICLE INFO ABSTRACT

It is very vital for suppliers and distributors to predict the deregulated electricity prices for creating their bidding strategies in the competitive market area. Pre requirement of succeeding in this field, accurate and suitable electricity tariff price forecasting tools are needed. In the presence of effective forecasting tools, taking the decisions of production, merchandising, maintenance and investment with the aim of maximizing the profits and benefits can be successively and effectively done. According to the electricity demand, there are four various electricity tariffs pricing in Turkey; monochromic, day, peak and night. The objective is find the best suitable tool for predicting the four pricing periods of electricity and produce short term forecasts (one year ahead-monthly). Our approach based on finding the best model, which ensures the smallest forecasting error measurements of; MAPE, MAD and MSD. We conduct a comparison of various forecasting approaches in total accounts for nine teen, at least all of those have different aspects of methodology. Our beginning step was doing forecasts for the year 2015. We validated and analyzed the performance of our best model and made comparisons to see how well the historical values of 2015 and forecasted data for that specific period matched. Results show that given the time-series data, the recommended models provided good forecasts. Second part of practice, we also include the year 2015, and compute all the models with the time series of January 2011 – December 2015. Again by choosing the best appropriate forecasting model, we conducted the forecast process and also analyze the impact of enhancing of time series periods (January, 2007 to December, 2015) to model that we used for forecasting process.

## 1. Introduction

Rapid growing of electricity consumption caused by nonstop growth in the world's population together with the desire for higher life standards. Facing with the demand problem for sustainability can be successively circumvented by efficient operation of energy market, affordable prices of energy production, prospering planning of energy investments and with the pertinent capacity decisions. Deregulated electricity price system is implemented in the industrial countries' electricity market. In that kind of market, the suppliers specify their electricity sale prices, whereas distributors give the need based bid. That competitive market provide qualified and continuous service with the low cost electricity prices to consumers. From the perspectival view of marketing company, with the competitive environment conditions explicated, making an short/medium/long term procurement agreement with the providers and other distributor companies, also agreements with the consumers, there is a need for forecasting beforehand the electrical tariff prices will be defined and emerged. In a similar manner, it is obvious that the production companies need that information for taking decisions on production, maintenance and investments. Private companies generally owns the sector more than public, because competitive markets have been introduced for wholesale trading business and retail markets gradually beat the local franchises. Typically the industry has been split up into separate companies for generation, transmission, local distribution and retail supply [1]. Successful and detailed modeling of prices become a very important aspect of risk management in the competitive market conditions.

With the limited numbers of local energy sources in Turkey, she has to import 65% of primary energy to meet the demand [2]. Turkey's electric energy consumption is expected to continue to grow rapidly at approximately 8% per year [3]. Within this scope decision takers and policy makers in Turkey have to develop methods for reducing foreign dependency and lead the electricity market economize by the reducing the electricity consumption and setting the tariff prices correctly in due to demand that will emerge.

Turkish electricity market, which is developing expeditiously as a result of growth of population, industrialization and urbanization, was controlled by the government. Within the last decade, intention of creating electricity market and transforming the monopolistic production to competitive market gave cause to legal reforms had been take place [4].

Prediction the electricity price is important for analyzing the projections for future. There can be fluctuations in the electricity demand, because of this situation electricity prices can vary in short-time periods. In Turkey, we see three periods in a day, namely day, peak and night; and the market is structured in according to day ahead pricing system. Main factors that influence daily electricity price are still under investigation. The studies in the literature, for predicting the electricity price, along with the historical price data, the indirect factors; heat, gale force, humidity, electricity demand, coil price, fuel price etc., that have effect on price can be used [5]. While forecasting with the systems that used those factors as inputs; future dated data are needed. Therefore, usages of those systems are limited.

For analyzing the historical data that related to current event, applying statistical knowledge to the forecasting models properly is very vital because the results will used for predicting future events. Time series comprised of a collection of specific data that have been collected over time with the continuous period of time. That collected data may already be in a daily, weekly, monthly, quarterly, or yearly format. Trend, seasonal effect cyclical and irregular effects are the four components of time series data [6]. Analyzing of a time series by the help of forecasting methods needs historical data. With the acknowledgement of the future fit data will resemble itself from the occurred past data.

The data needed in forecast models will be in short in supply when the market is fully privatized due to secrecy issues, consequently it is very important to have an appropriate model that predicts short-time periods with minimum error and less data; on the contrary with traditional time series methods that require large data.

Electricity price forecasting can be categorized into three different categories based on time horizons: short term (mainly one day ahead), medium-term (six months to one year) and long-term (one year and more).

The smallest time period for long term forecast is one year, and more than one year get the nod for long term forecasting. Long term forecasting can be essential for capacity expansion, investment of capital and profit analysis. Though the electricity market isn't stable because of the fluctuations in the demand and unlike load forecasting the electricity market itself has uncertainty characteristic in general with the factors of climate change, changing of daily life habits and industrial growth, , forecasting the electricity price forecasting is much more complex [7].

Various number of methods and models have been developed forecasting, especially for short term periods [8]. Stationary and non-stationary time series models can be used for short term predictions [9]. Auto-Regressive Integrated Moving Average (ARIMA) is used for forecasting both short-term and long-term periods [10].

There are several methods of statistical forecasting such as regression analysis, classical decomposition method, Box and Jenkins and smoothing techniques. All these different techniques give different accuracy. Accuracy of predicted model is determined by the error measurements. On no account, several factors such as prediction interval, prediction period, characteristic and size of time series affect the error measurements of various techniques [11].

In this research for forecasting monochromic, day, peak and night electrical tariff prices, we are interested in nine teen various approaches; which had got different aspects through others, like different seasonality characteristics; twelve and four, different kind of models while applying classical decomposition technique; with additive and multiplicative models, and based on different techniques; standard classical decomposition model, classical decomposition model with centering moving average method, regression equations, single, double and triple exponential (Holt Winters model) smoothing models and lastly ARIMA model. The most suitable forecasting method and the best choice of period were chosen by considering the smallest values of MAPE, MAD and MSD.

The remainder of the paper is structured as follows. Section 2 describes the various forecasting models that we used in that study and also gave the theoretical equations. Section 3 describes the best forecasting model, impact of the increasing the time period and one year future forecast values. Section 4 concluded the study in this paper, gave the results of our findings, fathomed out and emphasized our study.

## 2. Choosing a Forecasting Technique

### 2.1. Overview

Various questions can occur before choosing the right forecasting model for predicting the data you will compute. Some of those: why is a forecast needed, what are the characteristics of the available data, what time period will be chosen, and also what are the minimum data needed for that process, and how much accuracy will be needed…

In order to choose the right forecasting technique, we have to identify the nature of forecasting problem, choose the appropriate seasonal period that fits to the technique we will choose to study with, and at the end not to get high errors with the model is crucial. Historical patterns in the data is the most important aspect of selection of forecasting technique, so it's vital to identify and understand of patterns must have been done correctly.

### 2.2. Series Characteristic and Time Period

When the series seems to be stationary, that means the environment of series exist is stable, not changing a lot. In that kind of situation, forecast can be updated when the new data will be available. If the series is not stationary, we have to convert into stationary one by differencing. Techniques for stationary series are: naive, simple averaging, moving average and autoregressive moving average (Box Jenkins) methods.

Clear growth or decline in series, must sign us the being of trend characteristics in the series. Techniques for trending series: moving averages, autoregressive integrated moving average, Holt's linear exponential smoothing, exponential models, simple regression and growth curves models.

Repetition of the similar characteristic behavior continuously in the similar time of periods means seasonality. Techniques for seasonal data forecasting models: classical decomposition, multiple regression, Census x-12, Winter's exponential smoothing and ARIMA models.

Time period that we will choose has a major impact on selection of a forecasting technique. In the case of short and intermediate term forecasts; various techniques can be applied. As the time period increases, however, a number of the techniques become less. Regression models are very well fit to short, intermediate and long terms forecasting. Means, moving averages, classical decomposition and trend projections are appropriate for the short and intermediate periods. On behalf of periods with short horizontal time, exponential smoothing, trend projection, regression models and classical decomposition methods are desirable. Choosing best method that we will use in forecasting will be determined by comparing the error values of techniques, best method means the method with the smaller forecasting error.

### 2.3. Models in the Study

The particular forecasting techniques chosen for this study are comprised of; standard classical decomposition model (with seasonality is 12 or 4), classical decomposition model with centering moving averages method (with seasonality is 12 or 4), regression equations methodology (with seasonality is 12 or 4), exponential smoothing models ; single and double exponential smoothing models (with seasonality is 12 or 4) and Holt's Winter's method (with seasonality is 12 or 4) and ARIMA model methodology (with seasonality is 12 or 4).

**Table 1:** Our nine teen various approach for forecasting

| # | Description of forecasting approaches |
|---|---|
| 1 | Classical decomposition with multiplicative model with seasonality is 12 |
| 2 | Classical decomposition with multiplicative model with seasonality is 4 |
| 3 | Classical decomposition with additive model with seasonality is 12 |
| 4 | Classical decomposition with additive model with seasonality is 4 |
| 5 | Classical decomposition with centering moving averages with multiplicative model with seasonality is 12 |
| 6 | Classical decomposition with centering moving averages with multiplicative model with seasonality is 4 |
| 7 | Classical decomposition with centering moving averages with additive model with seasonality is 12 |
| 8 | Classical decomposition with centering moving averages with additive model with seasonality is 4 |
| 9 | Forecasting with regression equation with seasonality is 12 |
| 10 | Forecasting with regression equation with seasonality is 4 |
| 11 | Single exponential smoothing with seasonality is 12 |
| 12 | Single exponential smoothing with seasonality is 4 |
| 13 | Double exponential smoothing with seasonality is 12 with multiplicative model |
| 14 | Double exponential smoothing with additive model with seasonality is 12 |
| 15 | Double exponential smoothing with multiplicative model with seasonality is 4 |
| 16 | Double exponential smoothing with additive model with seasonality is 4 |
| 17 | Holt Winter's model with ideal coefficients with seasonality is 12 |
| 18 | Holt Winter's model with ideal coefficients with seasonality is 4 |
| 19 | ARIMA models |

### 2.3.1 Classical Decomposition Models

Decomposition process is comprised of identifying the component factors that influence each of the values in a series. Each component is identified as separately. With the projection use of each components, forecasting of future values make possible. With ignoring cyclical component, often case three components take into consideration in decomposition models; trend, seasonal and irregular components.

A model that treats the time series values as a sum of the components is called an additive components model, that model, observed value ($Y_t$), consists three components; seasonality component ($S_t$), trend component ($T_t$) and irregular component ($I_t$). Notation of additive composition model;

$$Y_t = T_t + I_t + S_t \qquad (1)$$

A model that treats the time series values as the product of the components is called a multiplicative components model, that model, observed value ($Y_t$), consists three components; seasonality component ($S_t$), trend component ($T_t$) and irregular component ($I_t$). Notation of multiplicative composition model;

$$Y_t = T_t \times I_t \times S_t \qquad (2)$$

### 2.3.2 Classical Decomposition with Centering Moving Average Model

In classical decomposition method, we obtained the seasonal indexes in order to forecast with the components, by help of the web site called Wessa.net. On the contrary with this method, we assigned equal weights to each observation by using centering moving average approach in order to calculate seasonal indexes and afterwards doing those indexes in forecasting. Notation of centering moving average model, $Y_t$ represents the actual value, $Y_{t+1}$ represents the forecasted value and k represents the number of terms in the moving average, is shown below.

$$Y_{t+1} = (Y_t + Y_{t-1} + \ldots + Y_{t-k+1}) / k \qquad (3)$$

### 2.3.3 Regression Equations

Many real life forecasting situations are not simple, so we can't use simple linear regression equation; in which the relationship between a single independent variable and a dependent variable is investigated. Regression models with more than one independent variable are called multiple regression models. Creating dummy variables will be called for when it is necessary to determine how a dependent variable is related to an independent variable. Our general regression equation form when the seasonality is 4 and 12 is shown below;

$$Y_t = C_0 + T_0 \cdot t + \beta_2 \cdot s_2 + \beta_3 \cdot s_3 + \beta_4 \cdot s_4 \quad (\text{when } s=4) \qquad (4)$$

$$Y_t = C_0 + T_0 \cdot t + \beta_2 \cdot s_2 + \beta_3 \cdot s_3 + \beta_4 \cdot s_4 + \beta_5 \cdot s_5 + \beta_6 \cdot s_6 + \beta_7 \cdot s_7 + B_8 \cdot s_8 + \beta_9 \cdot s_9 + \beta_{10} \cdot s_{10} + \beta_{11} \cdot s_{11} + \beta_{12} \cdot s_{12} \quad (\text{when } s=12) \qquad (5)$$

### 2.3.4 Exponential Smoothing Models

Exponential smoothing models include; single exponential smoothing, double exponential smoothing, Holt-Winter's smoothing. We all applied those models with seasonality is 4 and 12 individually. Exponential smoothing continuously revises an estimate in the light of more recent experiences. This method is based on smoothing historical values of a series in an exponentially decreasing manner. General notations for single exponential smoothing, double exponential smoothing and Holt-Winter's smoothing models are shown below with the smoothing constants; α, β and ⴗ. $T_t$ and $C_t$ represented the smoothed trend and constant value individually. $Y_t$ represented real value in that time period and $F_{t+1}$ symbolized forecasted future value and $F_t$ represented forecasted value for the time period of t.

#### 2.3.4.1 Single Exponential Smoothing Model

Single smoothing is used when no linear trend is present in the time series. Single exponential smoothing model is appropriate for short term forecasting.
Forecasting equation with single smoothing constant is shown below.

$$F_{t+1} = \alpha \cdot Y_t + (1-\alpha) \cdot F_t \qquad (6)$$

#### 2.3.4.2 Double Exponential Smoothing Model

When the time series has an increasing or decreasing trend, a modification to the single exponential smoothing model is used to account for trend. In double exponential smoothing model, a second smoothing constant, β, is included to account for the trend. Double exponential smoothing model is also appropriate for short term forecasting. Equations for double exponential smoothing model, p letter represents the forecasted period into the future, are shown below.

$$F_{t+1} = C_t + p \cdot T_t \qquad (7)$$

#### 2.3.4.3 Holt-Winter's Smoothing Model

In the case of data show both trend and seasonality, double exponential smoothing model will become useless. With the intention of handling seasonality, we have to add a third parameter, ⴗ. Each observation is the product of a non-seasonal value and a seasonal index for that particular period in that technique [12]. $S_t$ represents overall smoothing, $b_t$ represents trend smoothing, $I_t$ represents seasonal smoothing, L represents the length of periods and m represents the number of period that will be used in forecasting.

$$F_{t+m} = (S_t + m \cdot b_t) \cdot I_{t-L+m} \qquad (8)$$

### 2.3.5 ARIMA Models

Independent variables don't exist in the form of ARIMA models. Rather than, model uses iterative approach of identifying a possible model from a general class of models. The chosen is then compared with the historical data to see whether the model is accurate. In what extent that the chosen model is accurate dawn out by looking to residuals whether they are generally small and they are randomly distributed. If the specified model is not satisfactory, the process is repeated using a new model designed to improve on the original one. This iterative procedure continues until a satisfactory model is found. ARIMA models had been accepted as an effective tool for short term forecasting.

#### 2.3.5.1 Autoregressive Models

Autoregressive model has the appearance of a regression model with lagged values of the dependent variable in the independent variable positions. The symbol of Φ is related to constant level of the series and autoregressive models are suitable for stationary time series. A first order autoregressive model equation is shown below., where Φ represents the coefficients to be estimated and $Y_t$ and lagged ones represent the response variable at time t.

$$Y_t = \Phi_0 + \Phi_1 \cdot Y_{t-1} + \Phi_2 \cdot Y_{t-2} + \ldots + \Phi_p \cdot Y_{t-p} \qquad (9)$$

#### 2.3.5.2 Moving Average Models

Deviation of the response from its mean is a linear combination of current and past errors, that situation referred as moving average. We bared out that as time moves forward, the errors involved in that linear combination move forward as well. The general notation of the moving average model has the coefficients to be estimated: ω, the constant mean of the process: μ, the error term in that current time and also past errors, the response variable at time t: $Y_t$ and q as the number of past error terms. Equation of moving average model is shown below.

$$Y_t = \mu + \varepsilon_t - \omega_1 \cdot \varepsilon_{t-1} - \omega_2 \cdot \varepsilon_{t-2} - \ldots - \omega_q \cdot \varepsilon_{t-q} \quad (10)$$

### 2.3.5.3 Autoregressive Moving Average Models

In the case of moving average terms and autoregressive terms combine, we will have autoregressive moving average model with the p and q order, gives the order of autoregressive and moving average part correspondingly. The general equation of autoregressive moving average model is shown below.

$$Y_t = \Phi_0 + \Phi_1 \cdot Y_{t-1} + \Phi_2 \cdot Y_{t-2} + \ldots + \Phi_p \cdot Y_{t-p} + \varepsilon_t - \omega_1 \cdot \varepsilon_{t-1} - \omega_2 \cdot \varepsilon_{t-2} - \ldots - \omega_q \cdot \varepsilon_{t-q} \quad (11)$$

## 2.4. Measuring Forecast Error

### 2.4.1 Mean Absolute Deviation (MAD)

Mean Absolute Deviation (MAD) calculates the forecast accuracy by averaging the absolute values of the forecast errors.

$$MAD = \frac{1}{n} \sum_{t=1}^{n} |Y_t - Y\hat{}_t| \quad (12)$$

### 2.4.2 Mean Squared Error (MSE)

This method penalizes large forecasting errors, due to being of squared term in the equation.

$$MSE = \frac{1}{n} \sum_{t=1}^{n} (Y_t - Y\hat{}_t)^2 \quad (13)$$

### 2.4.3 Root Mean Squared Error (RMSE)

Like the method MSE, RMSE penalizes large errors, however it has identical units as the series being forecast so its magnitude is more easily can be interpreted.

$$RMSE = \sqrt{\frac{1}{n} \sum_{t=1}^{n} (Y_t - Y\hat{}_t)^2} \quad (14)$$

### 2.4.4 Mean Absolute Percentage Error (MAPE)

It will be more beneficial to calculate the forecast errors in terms of percentages rather than amounts in some cases. MAPE is calculated by calculating the absolute error in each period, dividing this by actual value for that period, and finding the average of absolute percentage errors. The result at the final will be multiplied by hundred, and that will allow us to express the error as percentage.

$$MAPE = \frac{1}{n} \sum_{t=1}^{n} \frac{|Y_t - Y\hat{}_t|}{|Y_t|} \quad (15)$$

### 2.4.5 Mean Percentage Error (MPE)

Sometimes we would want to know our forecasting method which we apply is biased, forecasting low or high. In those types of situations we use MPE. The rest we find is multiplied by 100 and expressed as a percentage. If the forecasting is unbiased, our MPE method result will be close to zero. If the result will be large negative percentage, the forecasting method is overestimating. If the result is large positive percentage, the forecasting method is underestimating.

$$MPE = \frac{1}{n} \sum_{t=1}^{n} \frac{(Y_t - Y\hat{}_t)}{Y_t} \quad (16)$$

## 2.5. Methodology Description

Firstly, with the time series period of 2011-2014, 19 different approaches for forecasting data were executed for monochromic, day, peak and night time electrical tariff prices and we found the error measures for all methods for all the different time periods of electricity pricing individually, and afterwards comparing of MAPE, MAD and MSD values needed to be done for choosing best appropriate forecasting method. The best appropriate method is the model which had the smallest error measurements of MAPE, MAD and MSD. After finding the best appropriate model for forecasting, we forecasted for one year ahead; 2015, hence forwards we compared the values forecasted with the real 2015 values. That step revealed the success criterion of out method we had chosen, method will smaller error measures, by letting us to see whether the forecasted 2015 year values were in what extend correct or not.

After bringing the success of our method to light, we extended the time period of series that we used in our forecasting model in order to see the impact of time period length, we took the time series period of 2007-2014. Again finding the best appropriate model for forecasting, we compared the MAPE, MAD and MSD values with the best model we had found previously with the time series period of 2011-2014. After doing the comparison of error measures between two different time period, we came to a conclusion and finally added the time series values up till today, the final time series period is between January 2007 – May 2016, and made forecast for one year ahead from May 2016.

### 2.5.1 Flowchart of Methodology

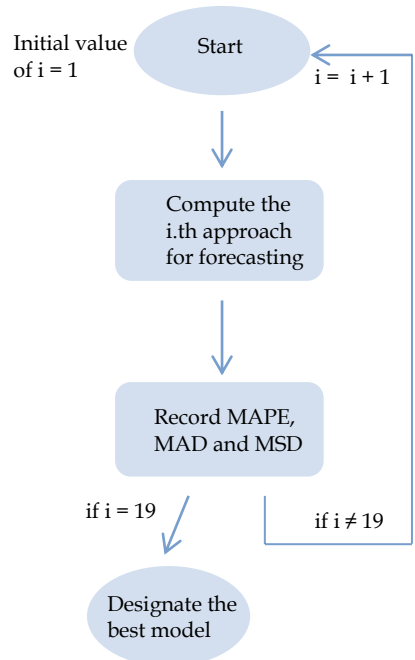

**Fig. 1**: Flowchart of methodology

## 3. Best forecasting methods for Monochromic, Day, Peak and Night Time Electrical Tariff Prices Forecasting within the time series of 2011-2014

Analyzing the series data of monochromic, day, peak and night time electrical tariff prices between 2011– 2014, which we provided in the supplementary document, with our series data we decided that there was a consistency over time with only little increase in amount, besides we saw the trend and also seasonality behavior. Examining the autocorrelation coefficients, between lags 1 to 12, neither of autocorrelation coefficients dropped to zero quickly and also none of them was close to zero. Series' data were not random. In Turkey the electrical tariff prices change by three month, without introducing that information to our model, we can expect little deviations on the validation of our forecasting values with the real values. We implemented all of our nine teen approaches which we defined in Table 1 before, and among all of them the three models with the seasonality of 12 came into prominence with the smallest error measures, the models were; Holt–Winter's smoothing model, double exponential smoothing model and regression equation model, and the error results were shown in Table 2 below. You can also reach all of the approaches' error measurements which provided in the supplementary document. On the basis of our results with the time series data of 2011 - 2014, both for the electrical tariff prices: monochromic, day, peak and night, Holt–Winter's smoothing model represented the most suitable forecasting models with the smallest error measures.

**Table 2:** Error measures of best three models.

| | | Monochromic | Day | Peak | Night |
|---|---|---|---|---|---|
| Holt – Winter's Smoothing Model s = 12 | MAPE | 1,04584 | 1,03415 | 1,10278 | 0,94636 |
| | MAD | 0,00274 | 0,00256 | 0,00425 | 0,00140 |
| | MSD | 3,15251 E-05 | 2,74256 E-05 | 7,82635 E-05 | 9,38699 E-05 |
| Double Exponential Smoothing Model s = 12 | MAPE | 1,69263 | 1,69190 | 1,69556 | 1,66888 |
| | MAD | 0,00452 | 0,00426 | 0,00669 | 0,00249 |
| | MSD | 7,14024 E-05 | 6,31615 E-05 | 0,00016 | 2,3164 E-05 |
| Regression Equation Model s = 12 | MAPE | 2,85416 | 2,82190 | 3,01593 | 3,23215 |
| | MAD | 0,00746 | 0,00694 | 0,01163 | 0,00464 |
| | MSD | 7,7079 E-05 | 6,82629 E-05 | 0,00017 | 2,76252 E-05 |

After finding the best forecasting model for each electrical tariff prices, performance validation had been done by comparing the historical and forecasted data for the same timeframe to see how our results for that period matched and correlated. As a result, we tried to see how well our forecasts tracked actual data. Such efforts reflected how these recommended models captured most of the characteristics of the time series data. Figure 2 showed that validation process for all of the electrical tariff prices; monochromic, day, peak and night for the year 2015 with comparing actual and fit values. Figure (2) shows the kinship level of the forecasted and real values for the monochromic, day, peak and night electrical tariff prices in Turkey for the year from January 2015 to December 2015. The forecasts and the actual did behave in the same manner, however, a careful examination of the figure shows for the first three months, the forecasted values were slightly below the actual values and then the forecasted values slightly brimmed over the real values. Our forecast technique worked on the crest of a wave; except incidental fit values to March of monochromic, day values of; April, October, November and December all the fit values were same in the manner 1 decimal places. For who want to enter more details, we also shared the numerical values of real and forecasted numerical values for the electrical tariff prices in Turkey for the year 2015 in our supplementary document. Estimation errors were needed to be shown also, remember our forecasted values were generated by our best model, Holt-Winter's smoothing, hereby we shared the approximation percentage errors in the Table 3 with; the negative sign expressed that our forecast was underestimated, controversy overestimated situation expressed with the positive sign in the approximation error value.

**Table 3:** 2015 Electrical tariff prices forecasts' approximation percentage errors.

| | Monochromic (%) | Day (%) | Peak (%) | Night (%) |
|---|---|---|---|---|
| January | -3,15 | -3,10 | -3,43 | -2,46 |
| February | -3,27 | -3,22 | -3,54 | 1,27 |
| March | -3,42 | -3,36 | -3,68 | -2,71 |
| April | 2,45 | 2,65 | 2,80 | 1,65 |
| May | 2,29 | 2,23 | 2,62 | 1,53 |
| June | 2,06 | 1,99 | 2,47 | 1,14 |
| July | 1,92 | 1,84 | 2,34 | 0,98 |
| August | 1,78 | 1,69 | 2,21 | 0,82 |
| September | 1,61 | 1,53 | 2,05 | 0,65 |
| October | 7,81 | 7,81 | 7,89 | 7,70 |
| November | 7,60 | 7,60 | 7,69 | 7,48 |
| December | 7,41 | 7,40 | 7,50 | 7,28 |

Between January – March, our approximation were underestimated as the negative sign was seen in the results, there were seemed to be little increment in the deviation amount in the latest three months; October, November and December, however that deviation amounts become important in the sensibility manner with the decimal places two or more. In summary, Holt-Winter's smoothing model could be accepted to performed well in decimal places of 1-2, in case of better forecasting results with more sensitivity measures in the manner of decimal places, we investigated the impact of extension of time series length, in that case we took the data between 2007 – 2014 for forecasting 2015 year fit values.

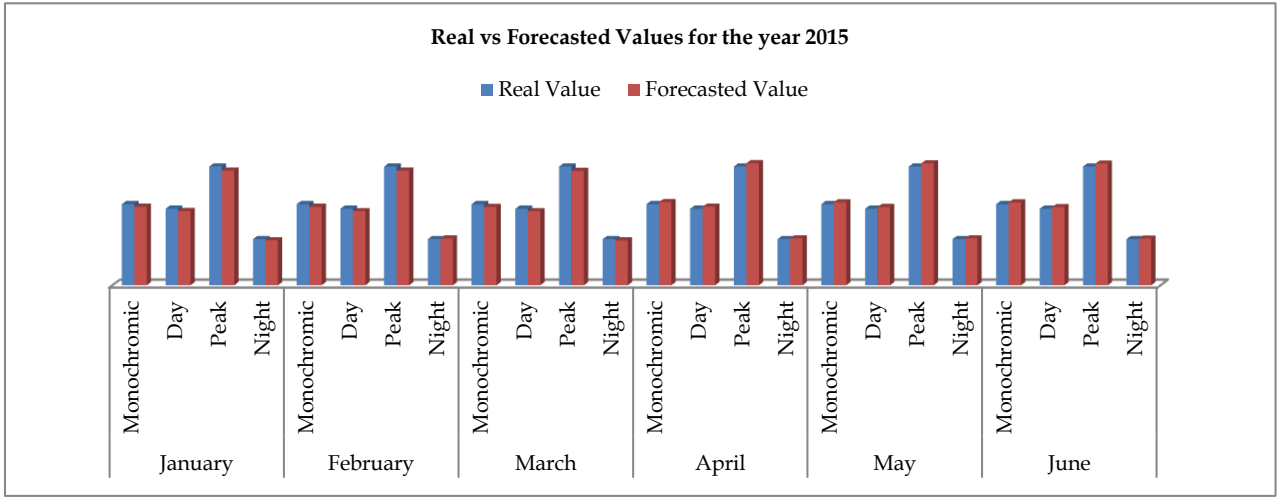

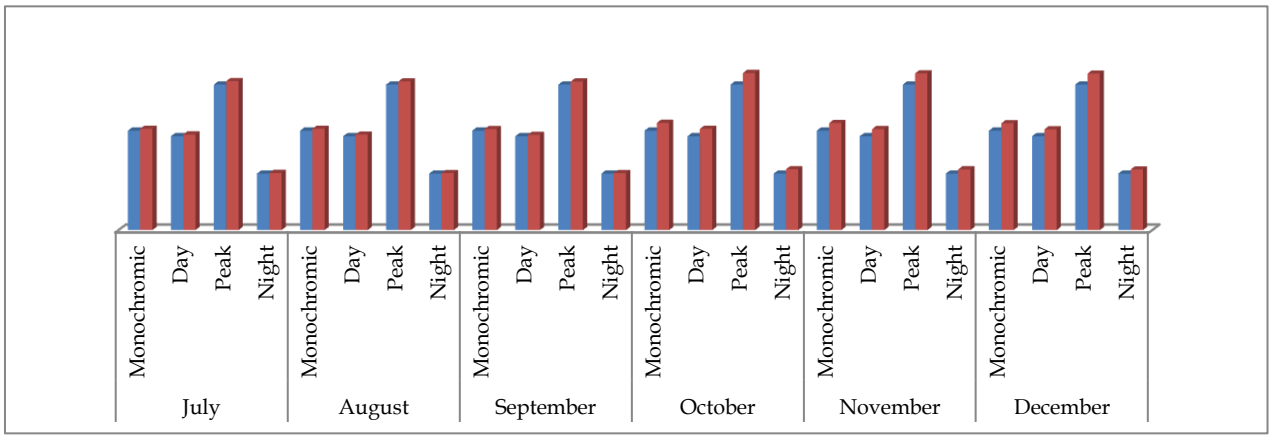

**Fig. 2**: Real vs Forecasted Values for the year 2015 for monochromic, day, peak and night electrical tariff prices in Turkey

### 3.1 Impact of Extending the Time Series Period

We implemented all of our nine teen approaches once again, and among all of them the three models shined out with the smallest error measures, the models were; Holt–Winter's smoothing and double exponential smoothing models with the seasonality was 12, and ARIMA models, the error results were shown in Table 4. Holt–Winter's and Double exponential smoothing models were two models that had smaller error measures than the rest of 16 models in the both case when the time series were taken 2011 – 2014 and also 2007 – 2014, however when we forecasted with the time series of 2007 – 2014, the best technique for forecasting became ARIMA model for all the various electrical tariffs; monochromic, day, peak and night.

The error results of our nine teen techniques and all the time series data between the 2007 – 2014 for the monochromic, day, peak and night electrical tariff prices for Turkey were given in our supplementary document. Analyzing the extension of seasonal period declared us that the best for forecasting model changed into ARIMA model. Also there was increment in the error measures of Holt-Winter's and double exponential smoothing model. Comparison of MAPE, MAD and MSD values of all the electrical tariff prices that had been forecasted by our best forecasting techniques with the time series period of 2011- 2014 and 2007 – 2014 was given also in the supplementary document. Additionally, forecasted values with the ARIMA models and real values for monochromic, day, peak and night electrical tariff prices for year 2015 also were given in the supplementary document. Validation step were carried out by comparing the corresponding best forecasting technique's approximation percentage error; with the time series period of 2011-2014 and 2007-2014 for all of the electrical tariff prices, the approximation percentage errors' comparison can be found in Figure (3). As it was seen from the Figure (3), the approximation percentage errors were lowered drastically. To put that situation more explicitly we had to put our expression into numerical manner; approximation percentage errors of our forecasts for all electrical tariffs with the time series period of 2011 – 2014; the months between January to March were averagely at % -2,7 - % 3,15 level; however with the time series period of 2007 – 2014, the approximation percentage error of months between January to March was % 0,59 - % 0,65. All of the other numerical values of percentage errors also were given in our supplementary document. Significant discrepancy was in the sign of the approximation percentage errors for all the one year; 2015. Whole of the errors were positive in sign when we put time series data between 2007 – 2014 to our model; albeit first three months' approximation percentage errors were negative in sign and the rest were positive in sign when we took the time series data of 2011 - 2014. Another analysis result showed that it's not only enough to look error measures for the finding best forecasting model, we need to evaluate both the error measures and the approximation percentage errors when comparing the different time series of period. The error measures of best forecasting method with time series of 2011-2014 were slightly smaller than the best model which had the time series of 2007-2014, however the approximation percentage errors showed our model worked much better with the time series of 2007 – 2014.

**Table 4:** Error measures of best three models when the time series period is 2007 – 2014.

| | | Mono chromic | Day | Peak | Night |
|---|---|---|---|---|---|
| Holt – Winter's Smoothing Model s = 12 | MAPE | 1,54134 | 1,53563 | 1,57967 | 2,00233 |
| | MAD | 0,00325 | 0,00305 | 0,00493 | 0,00239 |
| | MSD | 4,16305 E-05 | 3,64137 E-05 | 9,94564 E-05 | 1,79968 E-05 |
| Double Exponential Smoothing Model s = 12 | MAPE | 1,88821 | 1,83107 | 1,87624 | 2,06330 |
| | MAD | 0,00415 | 0,00374 | 0,00604 | 0,00249 |
| | MSD | 6,71492 E-05 | 5,34814 E-05 | 0,000141895 | 2,45896 E-05 |
| ARIMA model | MAPE | 1,22902 | 1,22134 | 1,26501 | 1,46693 |
| | MAD | 0,00277 | 0,00259 | 0,00422 | 0,00184 |
| | MSD | 5,49804 E-05 | 4,83688 E-05 | 0,000127408 | 2,39453 E-05 |

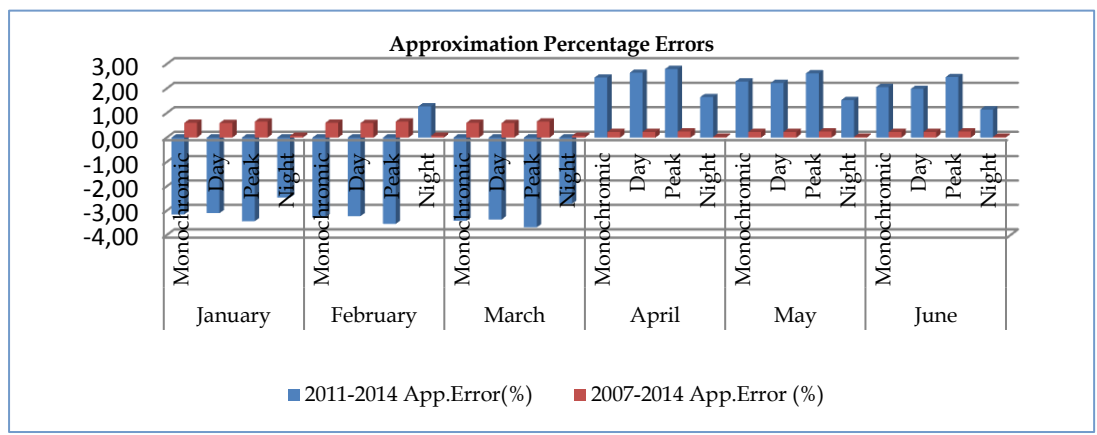

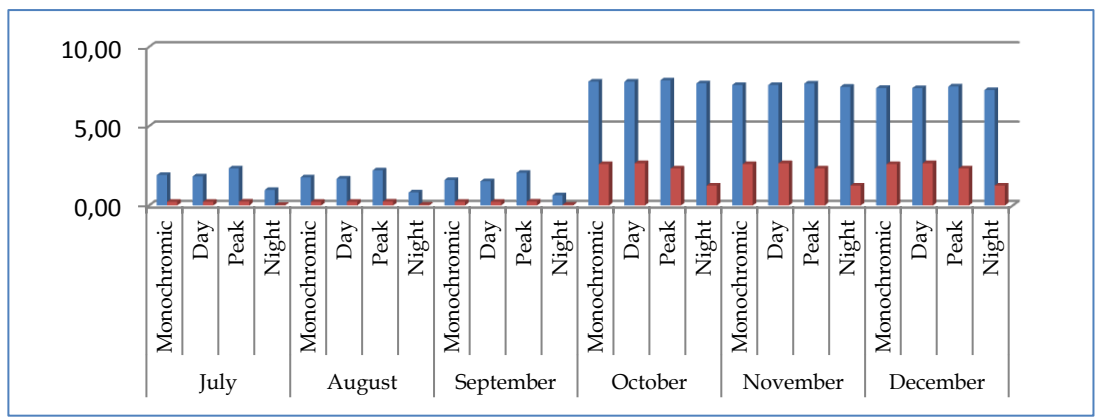

**Fig 3:** Comparison of approximation percentage errors

*3.2 Forecasting values for one year ahead*

We took the time series of period from January 2007 to June 2016 and made forecast for 1 year future; starting from July 2016. Best three models' error measures were given in Table 5. ARIMA (0,1,0) (0,0,1) model was to be found to be the best model for forecasting one year ahead electrical tariff prices in Turkey. We realized that the error measures of ARIMA model decreased (we only gave MAPE comparison with numerical) : for monochromic tariff MAPE measure; 1,22902 to 1,05792, for day tariff MAPE measure; 1,22134 to 1,09139, for peak tariff MAPE measure; 1,26501 to 1,12524, for night tariff MAPE measure; 1,46693 to 1,33358. That lessen situation was accepted to be stable, due to we had all encountered decrease in the error measures. Predicted values with ARIMA model for July 2016 to July 2017 can be found in our supplementary document.

**Table 5:** Error measures of best three models.

|   |   | Mono chromic | Day | Peak | Night |
|---|---|---|---|---|---|
| Holt – Winter's Smoothing Model s = 12 | MAPE | 1,41103 | 1,46192 | 1,48033 | 1,80495 |
|   | MAD | 0,00307 | 0,00305 | 0,00480 | 0,00228 |
|   | MSD | 3,86525 E -05 | 3,64243 E -05 | 9,52708 E -05 | 1,80295 E -05 |
| Double Exponential Smoothing Model s = 12 | MAPE | 1,63246 | 1,69036 | 1,69904 | 1,94033 |
|   | MAD | 0,00361 | 0,00358 | 0,00563 | 0,00247 |
|   | MSD | 5,29207 E -05 | 4,85419 E -05 | 0,00012 | 2,36791 E -05 |
| ARIMA model | MAPE | 1,05792 | 1,09139 | 1,12524 | 1,33358 |
|   | MAD | 0,00241 | 0,00239 | 0,00385 | 0,00174 |
|   | MSD | 4,73285 E -05 | 4,50851 E-05 | 0,00011 | 2,29849 E -05 |

**4. Conclusions**

Electrical tariff price forecasting has become one of the major research fields in recent years. The electric authorities require accurate forecasting tools for various time periods. Daily activity of electricity companies, consumers and supplier companies are in need of electricity tariff price short term forecasts. Accurate forecasting is crucial at optimization of the costs of electric energy and also for future planning and taking decisions.

There is no universal tool for price forecasting which can be useful for every market [14]. In the presented work a study of nineteen different forecasting approaches is done in order to find suitable electrical tariff prices predicting model for Turkey. The competitive market has created an increasing need to forecast accurate future prices with the intention of maximizing the profit. Forecasting with nineteen various approaches was compeller task, after all most suitable forecasting model was found. After finding the best model for forecasting, validating of our model's rightness needed to be done; we compared the forecasted 2015 year values and actual 2015 year values, and we figured out the approximation percentage errors. On the basis of our results, Holt-Winter's smoothing, double exponential smoothing and ARIMA model seemed to perform well. For the time period of 2011 to 2014, best forecasting model was Holt – Winter's smoothing model with the seasonality of 12; whereas for the period of 2007 to 2014; ARIMA (0,1,0) (0,0,1) model was found to be best with the smallest error measures. In the light of information; different time periods may have led to different results, we took two different time periods as we stated beforehand, and increasing the time series period caused decreasing in the error values and approximation percentage errors.

As we said before electrical tariff prices in Turkey change from three month to three month. Doubling the time series of our period (2007 – 2014) lead to our forecasting model did the character analysis of our time series data very well, mostly our model predicted correctly with two decimal places or three decimal places level. For all electrical tariff prices; all of the forecasted and real values when the time series period was between 2007 - 2014 were given in our supplementary document.

The whole process and methods used in this study could be great source for electric authorities, companies and consumers who want efficient and accurate forecasting tool for electricity price. The method and approaches we applied can be used in order to make better profit driven plans and taking low risk decisions. We strongly believe that our approach for finding the best forecasting model in the use of predicting Turkey's future electricity tariff price worked really very well. Not only Turkey's authorities can used that approach and model, also other countries' authorities can benefit from our work, we believe that our approach that had been applied can be universal one. Given the similar time series data and electricity price in other countries, the method and process used in this study can be worked reasonably well.